\documentclass[12pt]{article} \textwidth=17cm \textheight=22.5cm
\usepackage{psfig}
\def\beq{\begin{equation}} 
\def\eeq{\end{equation}} 
\def\bea{\begin{eqnarray}} 
\def\eea{\end{eqnarray}}
\def\bq{\begin{quote}} 
\def\eq{\end{quote}}
\def\nn{\nonumber}
\def\dd{\displaystyle}

\def\gappeq{\mathrel{\rlap {\raise.5ex\hbox{$>$}} {\lower.5ex\hbox{$\sim$}}}}

\def\lappeq{\mathrel{\rlap{\raise.5ex\hbox{$<$}} {\lower.5ex\hbox{$\sim$}}}}

\begin{document} \pagestyle{empty} \begin{flushright} {CERN-TH.2000-171\\DFPD 00/TH/34} \end{flushright}
\vspace*{5mm}
\begin{center} 
{\bf From Minimal to Realistic Supersymmetric SU(5) Grand Unification} \\ 
\vspace*{1cm}  
{\bf Guido Altarelli} \\ 
\vspace{0.3cm} 
Theoretical Physics Division, CERN \\ 
CH - 1211 Geneva 23 \\ 
\vspace{0.3cm} 
{\bf Ferruccio Feruglio and Isabella Masina} \\ 
\vspace{0.3cm}  
Universit\`a di Padova
\\
and
\\
I.N.F.N., Sezione di Padova, Padua, Italy\vspace{0.3cm} \\
\vspace*{2cm}   
{\bf Abstract} \\
\end{center} 
\vspace*{1cm} We construct and discuss a "realistic" example of SUSY 
$SU(5)$ GUT model,
with an additional $U(1)$ flavour symmetry, that is not
plagued by the need of large fine tunings, like those associated with
doublet-triplet splitting in the minimal model, and that 
leads to an acceptable phenomenology. This includes coupling unification
with a value of $\alpha_s(m_Z)$ in much better
agreement with the data than in the minimal version, an acceptable
hierarchical pattern for fermion masses and mixing angles,
also including neutrino masses and mixings, and a proton decay rate
compatible
with present limits (but the discovery of proton decay should be within
reach of the next generation of experiments). In the
neutrino sector the preferred solution is one with nearly maximal mixing both for
atmospheric and solar neutrinos. 

\vspace*{1cm}  \noindent  %\rule[.1in]{16.5cm}{.002in}

\noindent

\begin{flushleft} CERN-TH.2000-171 \\DFPD 00/TH/34 \\ June 2000 \end{flushleft} \vfill\eject %\pagestyle{empty}
%\clearpage\mbox{}\clearpage

\setcounter{page}{1} \pagestyle{plain}

%INSERT YOUR TEXT HERE

\section{Introduction}

The idea that all particle interactions merge into a unified theory \cite{uth}
at very high energies is so attractive that this
concept has become widely accepted by now. The quantitative success 
of coupling unification in Supersymmetric (SUSY)
Grand Unified Theories (GUT's) has added much support to this 
idea \cite{sgut}. 
The recent developments on neutrino oscillations \cite{osc},
pointing to lepton number violation at large scales, have further 
strengthened the general confidence.  However the
actual realization of this  idea is not precisely defined. On the one hand, 
at the Planck scale, $M_{Pl}$, the
unification of gravity with gauge interactions is a general 
property of superstring theories. On the other hand, in
simple GUT models gauge interactions are unified at a distinctly 
lower mass scale $M_{GUT}$ within the context of a
renormalizable gauge theory. But this neat separation between 
gauge unification and merging with gravity is not at all
granted. The gap between
$M_{GUT}$ and 
$M_{Pl}$ could be filled up by a number of threshold effects 
and several layers of additional states. Also, coupling
unification could be realized without gauge unification, 
as suggested in some versions of superstring theory or in
flipped $SU(5)$. 

Assuming gauge unification, minimal models of GUT's based on $SU(5)$, 
$SO(10)$, ... have been considered in detail. Also,
many articles have addressed particular aspects of GUT's models 
like proton decay, fermion masses and, recently,
neutrino masses and mixings. But minimal models are not plausible 
as they need a large amount of fine tuning and are
therefore highly unnatural (for example with respect to the 
doublet-triplet splitting problem or proton decay). Also,
analyses of particular aspects of GUT's often leave aside the 
problem of embedding the sector under discussion into a
consistent whole. So the problem arises of going beyond 
minimal toy models by formulating sufficiently realistic, not
unnecessarily complicated, relatively complete models that can 
serve as benchmarks to be compared with experiment. More
appropriately, instead of "realistic" we should say 
"not grossly unrealistic" because it is clear that many important
details cannot be sufficiently controlled and assumptions must be made.
The model we aim at should not rely on large
fine tunings and must lead to an acceptable phenomenology. 
This includes coupling unification with an acceptable value of
$\alpha_s(m_Z)$, given $\alpha$ and
$sin^2\theta_W$ at $m_Z$, compatibility with the more and more stringent 
bounds on proton decay \cite{pbounds, lastSK}, agreement with the observed 
fermion mass
spectrum, also considering neutrino masses and mixings and so on. 
The success or failure of the program of
constructing realistic models can decide whether or not 
a stage of gauge unification is a likely possibility. 

Prompted by recent neutrino oscillation data some new studies on realistic GUT models have appeared in the context of
$SO(10)$ 
or larger groups \cite{recentso10}. In the present paper we address the question whether the smallest SUSY $SU(5)$ symmetry group can
still be considered as a basis for a realistic GUT model. We indeed present an explicit example of a realistic $SU(5)$
model, which uses a $U(1)$ flavour symmetry as a crucial ingredient. In principle the flavour symmetry could be either
global or local. We tentatively assume here that the flavour symmetry is global. This is more in the spirit of GUT's
in the sense that all gauge symmetries are unified. The associated Goldstone boson receives a mass from the anomaly.
Such a pseudo-Goldstone boson can be phenomenologically acceptable in view of the existing limits on axion-like
particles \cite{gkr}. 

In this model the doublet-triplet splitting problem is solved by the missing partner mechanism \cite{mdm} 
stabilized by
the flavour symmetry against the occurrence of doublet mass lifting due to non renormalizable operators. Relatively large
representations (50,
$\overline{50}$, 75) have to be introduced for this purpose. A good effect of this proliferation of states is that the value
of $\alpha_s(m_Z)$ obtained from coupling unification in the next to the leading order perturbative approximation
receives important negative corrections from threshold effects near the GUT scale. As a result, the central value
changes from 
$\alpha_s(m_Z)\approx 0.130$ in minimal SUSY $SU(5)$ down to $\alpha_s(m_Z)\approx 0.116$, in better agreement with
observation \cite{yam,bag,cla}. 
The same $U(1)$ flavour symmetry that stabilizes the missing partner mechanism is used to explain the
hierarchical structure of fermion masses. In the neutrino sector, the mass matrices already proposed in a previous paper
by two of us are reproduced \cite{af23}. 
The large atmospheric neutrino mixing is due to a large left-handed mixing in the lepton
sector that corresponds to a large right-handed mixing in the down quark sector. In the present particular version
maximal mixing also for solar neutrinos is preferred. A possibly problematic feature of the model is that, beyond the
unification point, when all the states participate in the running, the asymptotic freedom of $SU(5)$ is destroyed by the
large number of matter fields. As a consequence, the coupling increases very fast and the theory becomes non
perturbative below
$M_{Pl}$. In the past models similar to ours have been considered, but were discarded just because they contain many
additional states and tend to become non perturbative between
$M_{GUT}$ and
$M_{Pl}$. We instead argue that these features are not necessarily bad. While the predictivity of the theory is reduced
because of non renormalizable operators that are only suppressed by powers of $M_{GUT}/\Lambda$ with $\Lambda<M_{Pl}$, still
these corrections could explain the distortions of the mass spectrum with respect to the minimal model, the suppression
of proton decay and so on. However, it is certainly true that also in this case, as for any other known realistic model,
the resulting construction is considerably more complicated than in the corresponding minimal model.

\section{The Model}

The symmetry of the model is SUSY $SU(5)\otimes U(1)$. We call $Q$ the charge associated with $U(1)$. The superpotential
of the model has three parts:
\beq 
w=w_1+w_2+w_3~~~~. \label{1}
\eeq 
The $w_1$ term only contains the field $Y$ of the representation 75 of $SU(5)$ with $Q=0$:
\beq w_1=c_1~Y^3+M_Y~Y^2~~~~~.\label{2}
\eeq The effect of $w_1$ is to provide $Y$ with a vev of order $M_Y/c_1\approx M_Y\approx M_{GUT}$ and to give a mass to all
physical components of $Y$, i.e. those that are not absorbed by the Higgs mechanism (the 75 uniquely 
breaks $SU(5)$ down to $SU(3)\otimes SU(2)\otimes U(1)$).

The $w_2$ term induces the doublet-triplet splitting:
\beq w_2=c_2~HY H_{50}~+~c_3\bar HYH_{\overline{50}}~+~c_4~H_{50}{H_{\overline{50}}}X~~~.\label{3}
\eeq Here $H$ and $\bar{H}$ are the usual pentaplets of Higgs fields, except that now they carry non opposite $Q$ charges. It is not restrictive to take $c_2$, $c_3$ and $c_4$ real and positive.
The renormalizable couplings that appear in $w_2$ are the most general allowed by the $SU(5)$ and $Q$ assignments of the
fields in eq. (\ref{3}) which are given as follows:
%\beq
%\begin{array}{ccccccc}
%{\rm field}& Y& H& \bar{H}& H_{50}& {H_{\overline{50}}}& X\\
%SU(5)& 75& 5& \bar{5}& 50 &\bar{50} & 1\\
%Q& 0& -q& q-1& q& 1-q& -1 
%\end{array}
%\label{4}
%\eeq

\bea {\rm field}~~~~~~~~~Y~~~~~~~~~H~~~~~~~~~\bar H~~~~~~~~~H_{50}~~~~~~~~~{H_{\overline{50}}}~~~~~~~~~X \nonumber\\
SU(5)~~~~~~~~~75~~~~~~~~~5~~~~~~~~~~\bar 5~~~~~~~~~~50~~~~~~~~~~~\bar{50}~~~~~~~~~~~1\nonumber\\
~~~~Q~~~~~~~~~0~~~~~~~~-q~~~~~~~~~q-1~~~~~q~~~~~~~~~~1-q~~~~~~-1\label{4}
\eea  
The value of $q$ will be specified later. At the minimum of the potential, in the limit of unbroken SUSY, the vevs
of the fields $H$, $\bar H$,
$H_{50}$ and
${H_{\overline{50}}}$ all vanish, while the $X$ vev remains undetermined. 
As we shall see, when SUSY is softly broken the
light doublets in
$H$ and $\bar H$ acquire a small vev while the $X$ vev will be fixed near the cut-off 
$\Lambda$, of the order of the scale
between $M_{GUT}$ and $M_{Pl}$ where the theory becomes strongly interacting (we shall see that we estimate this scale
at around $10-20~ M_{GUT}$, large enough that the approximation of neglecting terms of order $M_{GUT}/\Lambda$ is not
unreasonable). The missing partner mechanism to solve the doublet-triplet splitting problem occurs because the 50
contains a $(\bar 3,1)$, i.e. a coloured antitriplet, $SU(2)$ singlet (of electric charge 1/3) but no colourless doublet
(1,2). The $U(1)$ flavour symmetry protects the doublet Higgs to take mass from radiative corrections because no $H\bar H$
mass term is allowed. Also no non renormalizable terms of the form $H\bar H Y^m X^n$ $(m,n\ge 0)$ are possible, because $X$ has a
negative $Q$ charge. This version of the missing partner mechanism 
was introduced in ref. \cite{ber} and overcomes the
observation in ref. \cite{ran} that, in general, non renormalizable interactions spoil the mechanism. The Higgs colour
triplets mix with the analogous states in the 50 and the resulting mass matrix is of the see-saw form:  
\beq \hat{m}_T= 
\left[\matrix{ 0&c_2\langle Y\rangle\cr c_3\langle Y\rangle&c_4\langle X\rangle    } 
\right]~~~~~. 
\label{5}
\eeq
Defining $m_{\phi}=c_4\langle X\rangle$ the eigenvalues of the matrix $\hat{m}_T \hat{m}_T^\dagger$ are
the squares of:
\beq
m_{T1,2}=\frac{1}{2}\left[
\sqrt{m_{\phi}^2+(c_2+c_3)^2\langle Y\rangle^2}\pm \sqrt{m_{\phi}^2+(c_2-c_3)^2\langle Y\rangle^2}\right]~~~.
\label{aut}
\eeq
Note that $m_{T1}m_{T2}=c_2 c_3 \langle Y\rangle^2$. 
The effective mass that enters in the dimension 5 operators with $|\Delta B=1|$ is 
\beq
m_T=\frac{{{m_T}_1} {{m_T}_2}}{m_{\phi}}=\frac{c_2c_3}{c_4} \frac{\langle Y\rangle^2}{\langle X\rangle}\label{mT}~~~.
\eeq

The $w_3$ term contains the Yukawa interactions of the quark and lepton fields $\Psi_{10}$, $\Psi_{\bar{5}}$ and
$\Psi_1$, transforming as the representations 10, $\bar 5$ and 1 of SU(5) respectively. We assume an exact 
$R$-parity discrete symmetry under which $\Psi_{10}$, $\Psi_{\bar{5}}$ and $\Psi_1$ are odd whereas
$H$, $\bar H$, $H_{50}$ and ${H_{\overline{50}}}$ are even. The $w_3$ term is symbolically given by
\bea 
w_3&=&\Psi_{10}G_u(X,Y)\Psi_{10}H~+~\Psi_{10}G_d(X,Y)\Psi_{\bar 5}\bar H~+~
\Psi_{\bar 5}G_{\nu}(X,Y)\Psi_{1}H\nn\\
&+&M\Psi_{1}G_M(X,Y)\Psi_{1}~+~\Psi_{10}{G_{\overline{50}}}(X,Y)\Psi_{10} H_{{\overline{50}}}~~~.
\label{w3}
\eea 
The Yukawa matrices $G_u$, $G_d$, $G_{\nu}$, $G_M$ and ${G_{\overline{50}}}$ depend on $X$ and $Y$ and the associated 
mass matrices on their
vevs. The last term does not contribute to the mass matrices because of the vanishing vev of ${H_{\overline{50}}}$, but
is important for proton decay. The pattern of fermion masses is determined by the $U(1)$ flavour symmetry that fixes the
powers of
$\lambda\equiv\langle X\rangle/\Lambda$ for each entry of the mass matrices.  In fact $X$ is the only field with non vanishing $Q$ that takes
a vev. The powers of $\lambda$ in the mass terms are fixed by the $Q$ charges of the matter field $\Psi$ and of the
Higgs fields $H$ and $\bar H$. We can then specify the charge $q$ that appears in (\ref{4}) and the $Q$ charges of the
matter fields $\Psi$ in order to obtain realistic textures for the fermion masses. We choose $q=2$, so that we have from
the table in (\ref{4}):
\beq Q(H)=-2~~{\rm and}~~Q(\bar H)=1~~~,\label{7}  
\eeq and, for matter fields
\beq 
Q(\Psi_{10})=(4,3,1)~~~,~~~~~~~Q(\Psi_{\bar 5})=(4,2,2)~~~,~~~~~~~Q(\Psi_{1})=(1,-1,0)~~~.\label{8} 
\eeq The Yukawa mass matrices are of the form:
\beq G_r(\langle X\rangle,\langle Y\rangle)_{ij}~=~\lambda^{n_{ij}}G_r(\langle Y\rangle)_{ij},~~~~~~~~{r=u, d, \nu, M}
~~~. \label{9}
\eeq We expand $G_r(\langle Y\rangle)_{ij}$ in powers of $\langle Y\rangle$ and consider the lowest order term at first. Taking $G_r(0)_{ij}$ of
order 1 and $n_{ij}$ as dictated by the above charge assignments we obtain \footnote{
In our convention Dirac mass terms are given by $L^T m R^*$ and the light neutrinos effective mass matrix is
$m_\nu m^{-1}_{maj}m_\nu^T$.}: 
\beq
\begin{array}{lr}
m_u~=~ 
\frac{1}{\sqrt{2}}\left[\matrix{
\lambda^6&\lambda^5&\lambda^3\cr
\lambda^5&\lambda^4&\lambda^2\cr
\lambda^3&\lambda^2&1    } 
\right]~v_u~~~~,& 
~~~~~~~~~~m_d~=~m_e^T~=~\frac{1}{\sqrt{2}} 
\left[\matrix{
\lambda^5&\lambda^3&\lambda^3\cr
\lambda^4&\lambda^2&\lambda^2\cr
\lambda^2 & 1 & 1    } 
\right]~v_d\lambda^4~~~~~,\\
& \\
m_{\nu}~=~\frac{1}{\sqrt{2}}
\left[\matrix{
\lambda^3&\lambda&\lambda^2\cr
\lambda&0&1\cr
\lambda&0&1    } 
\right]~v_u~~~~,&
m_{maj}~=~ 
\left[\matrix{
\lambda^2&1&\lambda\cr 1&0&0\cr
\lambda&0&1    } 
\right]~M~~~~. 
\label{Mnu}
\end{array}
\eeq
For a correct first approximation of the observed spectrum we need $\lambda\approx\lambda_C\approx 0.22$,
$\lambda_C$ being the Cabibbo angle. These mass matrices closely match those of ref. \cite{af23}, with two important
special features. First, we have here that
$\tan\beta=v_u/v_d\approx m_t/m_b\lambda^4$, which is small. The factor $\lambda^4$ is obtained as a consequence of the
Higgs and matter fields charges Q, while in ref. \cite{af23} the $H$ and $\bar H$ charges were taken as zero. We recall
that a value of $\tan\beta$ near 1 is an advantage for suppressing proton decay.  
A small range of $\tan\beta$ around one is currently disfavored by the negative results of the SUSY Higgs
search at LEP \cite{s2k}. Of course we could easily avoid this range, if necessary.
Second, the zero entries in the mass
matrices of the neutrino sector occur because the negatively $Q$-charged $X$ field has no counterpart with positive
$Q$-charge. Neglected small effects could partially fill up the zeroes. As explained in ref. \cite{af23} these zeroes lead
to near maximal mixing also for solar neutrinos. 

A problematic aspect of this zeroth order approximation to the mass
matrices is the relation $m_d=m_e^T$. This equality is good as an order of magnitude relation because relates large
left-handed mixings for leptons to large right handed mixings for 
down quarks \cite{af23,largem}. However the implied equalities
$m_b/m_{\tau}=m_s/m_{\mu}=m_d/m_e=1$ are good only for the third generation while need to be corrected by factors of 3
for the first two generations. The necessary corrective terms can arise from the neglected terms in the expansion in
$\langle Y\rangle$ of $G_r(\langle Y\rangle)_{ij}$ \cite{elg}. The higher order terms correspond to non renormalizable operators with the insertion of $n$
factors of the 75, which break the transposition relation between $m_d$ and $m_e$. For this purpose we would like the
expansion parameter $\langle Y\rangle/\Lambda$ to be not too small in order to naturally provide the required factors of 3.
We will present in the following explicit examples of parameter choices that lead to a realistic spectrum without
unacceptable fine tuning. 

The breaking of SUSY fixes a large vev for the field $X$ by removing the corresponding flat direction, gives
masses to s-partners, provides a small mass to the Higgs doublet and introduces 
a $\mu$ term. Up to coefficients 
of order one we can write down the terms that break SUSY softly:
\beq
-{\cal L}_{soft}~=~m^2 |x|^2+m^2|c|^2~+~m^2|\bar c|^2~+m(c \bar c x~+~h.c.)~+.....\label{14}
\eeq
where $c$, $\bar c$ and $x$ denote the scalar components of $H_{50}$, ${H_{\overline{50}}}$ and $X$ respectively, and dots stand for the remaining soft breaking terms, including mass terms for the scalar
components of the Higgs doublet fields
\footnote{To find the minima of the scalar
potential it is not restrictive to set to zero the imaginary part of $x$, which will be understood
in the remaining part of this section.}.
In the SUSY
limit and neglecting the mixing between the $(50,{\overline{50}})$ and the $(5,\bar{5})$ sectors, 
fermions and scalars in the $50$ and 
${\overline{50}}$ have a common squared mass $c^2_4 x^2$. 
When SUSY is broken by the soft terms for
each fermion of mass $c_4 x$ there are two bosons of squared masses $c^2_4 x^2+m^2\pm mx$. 
The $x$ terms in the scalar potential at one
loop accuracy are given by:
\bea
V&= &m^2 x^2+\frac{50}{64\pi^2}\left\{2(c^2_4 x^2+m^2+mx)^2 \left[\log\left({\frac{c^2_4 x^2+m^2+mx}{\Lambda^2}}\right)
-3/2\right]\right.\nn\\
&+&2(c^2_4 x^2+m^2-mx)^2
\left[\log\left({\frac{c^2_4 x^2+m^2-mx}{\Lambda^2}}\right)-3/2\right]\nn\\
&-&\left.4c^4_4 x^4\left[\log\left({\frac{c^2_4 x^2}
{\Lambda^2}}\right)-3/2\right]\right\}
\label{l15}
\eea
A numerical study of this potential in the limit $m\ll\Lambda$ and for $c_4$ of order one,
shows that the minimum is at $x$ of order $\Lambda$ (somewhat smaller than $\Lambda$ but close to it).
The expression for $V$ in eq. (\ref{l15}) provides a good approximation of the scalar potential
only in a small region around $\Lambda$. Outside this region the perturbative approximation will break 
down. We take our numerical analysis as an indication that the minimum occurs near $\Lambda$ and 
we assume that the true minimum occurs at $x=\langle X\rangle=0.25 \Lambda$,
that is $\lambda=0.25$. 
The complex field $x$ describes two physical scalar particles. The one associated to the real part of $x$
has a mass of order $m$ and couplings to the ordinary fermions suppressed by $1/\Lambda$.
The particle associated to the imaginary part of $x$ is massless only in the tree approximation. 
Due to the anomaly of the related $U(1)_Q$ current the particle acquires a mass of order
$f_\pi m_\pi/\langle X\rangle$. Cosmological bounds on $\langle X\rangle$ have been recently
reconsidered in ref. \cite{gkr} where it has been observed that $\langle X\rangle$ of the
order of the grand unification scale is not in conflict with observational data.

An alternative possibility is to assume that the $U(1)$ symmetry is local \cite{ber}. In this
case the supersymmetric action contains a Fayet-Iliopoulos term and the associated D-term in the 
scalar potential provides a large vev for $x$, of the order of the cut-off scale $\Lambda$.

A $\mu$ term for the fields $H$ and $\bar H$ of the appropriate order of magnitude can be generated according to the
Giudice-Masiero mechanism \cite{giu}. Assume that the breaking of SUSY is induced by the $\theta^2$ component of a chiral 
(effective) superfield $S$, singlet under $SU(5)$ with $Q=0$. In the Kahler potential a term of the form
\beq
K~=~\frac{S^\dagger X^\dagger H \bar H}{\Lambda^2}+{\rm h.c.}\label{16}
\eeq
is allowed. The vevs of $S$ and $X$, $\langle S\rangle\sim\theta^2 m M_{Pl}$, 
$\langle X\rangle=\lambda\Lambda$ lead to an equivalent term in the
superpotential of the desired form $\mu H \bar H$ with $\mu\sim \lambda m M_{Pl}/\Lambda$ which can be considered 
of the right order. 

\section{Coupling Unification}

It is well known that in the minimal version of SUSY SU(5) the central value of $\alpha_s(m_Z)$ required by the
constraint of coupling unification is somewhat large: 
$\alpha_s(m_Z)\approx 0.13$ \cite{yam,bag,cla,lp,cpp}. 
In the model discussed here, where
the doublet triplet splitting problem is solved by introducing the $SU(5)$ representations $50$, ${\overline{50}}$ and $75$, the
central value of $\alpha_s(m_Z)$ is modified by threshold corrections near $M_{GUT}$ which bring the central value down
by a substantial amount so that the final result can
 be in much better agreement with the observed value. As discussed in
refs. \cite{yam,bag}, this remarkable result arises because the $24$ of the minimal model is replaced by the $75$. The mass splittings
inside these representations are dictated by the group embedding of the singlet under
$SU(3)\otimes SU(2)
\otimes U(1)$ that breaks SU(5). The difference in the threshold contributions from the $24$ and the $75$ has the right
sign and amount to bring $\alpha_s(m_Z)$ down even below the observed value. The right value can then be
obtained by moving $m_{SUSY}$ and $m_T$ in a reasonable range. The difference in the favored value of $m_T$
with respect to the minimal model in order to reproduce the observed value of $\alpha_s(m_Z)$ goes in the right
direction to also considerably alleviate the potential problems from the bounds on proton decay. Note that there are no
additional threshold corrections from the
$50$ and
${\overline{50}}$ representations because the mass of these states arise from the coupling to the field $X$ which 
is an $SU(5)$ singlet. Thus there are no mass splittings inside the
$50$ and the
${\overline{50}}$ and no threshold contributions. While there is no effect on the value of $\alpha_s(m_Z)$ the presence of the
states in the
$50$ and
${\overline{50}}$ representations affects the value of the unified coupling at $M_{GUT}$ and also spoils the asymptotic
freedom of SU(5) beyond $M_{GUT}$. We find it suggestive that the solution of the doublet triplet splitting problem in
terms of the $50$, ${\overline{50}}$ and $75$ representations  automatically leads to improve the prediction of $\alpha_s(m_Z)$ and
at the same time relaxes the constraints from proton decay.  We now
discuss this issue in more detail.

Defining in the ${\overline{MS}}$ scheme $\alpha \equiv \alpha_{QED}(m_Z)$, $\sin^2{\theta}\equiv \sin^2{\theta}(m_Z)$,
$a_3\equiv \alpha_s(m_Z)$, $a_5\equiv \alpha_5(M_{GUT})$, from the one-loop renormalization group
evolution of the couplings in the MSSM with only two light Higgs doublets we have:
\bea
a_{3}^{(0)}=\frac{28\alpha}{60\sin^2{\theta}-12}\nonumber\\
a_{5}^{(0)}=\frac{28\alpha}{36\sin^2{\theta}-3}\nonumber\\
\log{(M^{(0)}_{GUT}/m_Z)}=\frac{\pi(3-8\sin^2{\theta})}{14 \alpha}\label{17}
\eea
where by $a^{(0)}$ we mean the quantity $a$ in the leading log approximation (LO). 
The LO results are the same as in
minimal SUSY $SU(5)$. For $\alpha^{-1}=127.934$ and $\sin^2{\theta}=0.231$ one 
obtains $a_{3}^{(0)}=0.118$ and 
$M^{(0)}_{GUT}=2.1~10^{16}$ GeV.

To go beyond the LO one must include two-loop effects in the running of gauge couplings, threshold
effects at the scale $m_{SUSY}$, close to the electroweak scale, and threshold effects at the large scale 
$M_{GUT}$. The unification scale $M_{GUT}$ is not univocally defined beyond LO and here 
we choose to identify it with the mass of the $SU(5)$ superheavy gauge bosons. The
results derived in ref. \cite{yam} can be written in the following form:
\bea
a_3&=&\frac{a_{3}^{(0)}}{(1+a_{3}^{(0)}\delta)}\nonumber\\
\delta&=&k+\frac{1}{2\pi}\log\frac{m_{SUSY}}{m_Z}-
\frac{3}{5\pi}\log\frac{m_T}{M_{GUT}^{(0)}} ~~~~.\label{18}
\eea  
In this expression for $\delta$ the logarithmic term with $m_{SUSY}$ comes from particles with masses near $m_{SUSY}$.
Similarly, the logarithmic term with $m_T$ arises from particles with masses of order $m_T\approx M_{GUT}$. The definition
of $m_T$ is the mass of Higgs colour triplets in the minimal model or the effective mass, defined in eq. (\ref{mT}), that, 
in the realistic model, plays the same role for proton decay \cite{yam,bag}. The term indicated with
$k=k(2)+k(SUSY)+k(M_{GUT})$ contains the contribution of two-loop diagrams to the running couplings,
$k(2)$, the threshold contribution of states near
$m_{SUSY}$, $k(SUSY)$, and the threshold contribution from states near $M_{GUT}$, $k(M_{GUT})$. The threshold contributions
would vanish if all states had the mass $m_{SUSY}$ or $M_{GUT}$ 
so that only mass splittings contribute to
$k(SUSY)$ and $k(M_{GUT})$. The values of $k(2)$ and of $k(SUSY)$ are essentially the same in the minimal and the realistic
model: typical values are
$k(2)=-0.733$ and $k(SUSY)=-0.510$. 
The value for $k(SUSY)=-0.510$ corresponds to the representative spectrum displayed in Table 1.
The value of $k(M_{GUT})$ is practically zero for the $24$ of the minimal model while
we have $k(M_{GUT})=1.857$ for the $75$ of the realistic model. Thus we obtain:
\bea
k=-0.733-0.510&=&-1.243~~~~~~~~\rm{minimal~ model}\nonumber\\
k=-0.733-0.510+1.857&=&0.614~~~~~~~~~~~\rm{realistic~ model}\label{k}
\eea
This difference is very important and makes the comparison with experiment of the predicted value
of $\alpha_s(m_Z)$ much more favorable in the case of the realistic model. In fact for $k$ large and negative as in the
minimal model we need to take $m_{SUSY}$ as large as
possible and $m_T$ as small as possible. But the smaller $m_T$ the faster is proton decay. The best compromise is
something like $m_{SUSY}\approx 1$ TeV and $m_T\approx M_{GUT}^{(0)}$, which 
leads to $\alpha_s(m_Z)\approx 0.13$ which is still
rather large and proton decay is dangerously fast. This is to be confronted with the case of the realistic model where $k$
is instead positive and large enough to drag $\alpha_s(m_Z)$ below the observed value. We now prefer $m_T$ to be larger
than $M_{GUT}^{(0)}$ by typically a factor of 20-30, which means a factor of 400-1000 of suppression for the proton decay rate with respect
to the minimal model. For example, for $m_{SUSY}\approx0.25$ TeV and $m_T\approx 6\cdot 10^{17}$ GeV we obtain 
$\alpha_s(m_Z)\approx 0.116$ which is acceptable.
The predictions for $a_3$ versus $(m_{SUSY},m_T)$ in the 
minimal and the missing doublet models
are shown in fig. 1.
\\[0.1cm]
{\begin{center}
\begin{tabular}{|c|c|}   
\hline
& \\                         
sparticle & mass$^2$\\ 
& \\
\hline& \\
gluinos & $(2.7 m_{1/2})^2$ \\& \\
\hline& \\
winos & $(0.8 m_{1/2})^2$\\& \\ 
\hline& \\
higgsinos & $\mu^2$\\& \\
\hline& \\
extra Higgses & $m_H^2$\\& \\
\hline& \\
squarks & $m_0^2+6 m^2_{1/2}$\\& \\
\hline& \\
(sleptons)$_L$ & $m_0^2+0.5 m^2_{1/2}$\\& \\
\hline& \\
(sleptons)$_R$ & $m_0^2+0.15 m^2_{1/2}$\\& \\
\hline
\end{tabular} 
\end{center}}
\vspace{3mm}
Table 1. Representative SUSY spectrum. $SU(2)\otimes U(1)$ breaking effects are neglected.
The additional freedom related to the parameters $m_0$, $m_{1/2}$, $\mu$, $m_H$ is here fixed 
by choosing $0.8 m_0=0.8 m_{1/2}=2 \mu=m_H$ and taking as a definition
$m_{SUSY}\equiv m_H$,
so that all particle masses can be expressed in term of $m_{SUSY}$. This parametrization
leads to $k(SUSY)=-0.510$.

\begin{figure}[p]
\centerline{\psfig{file=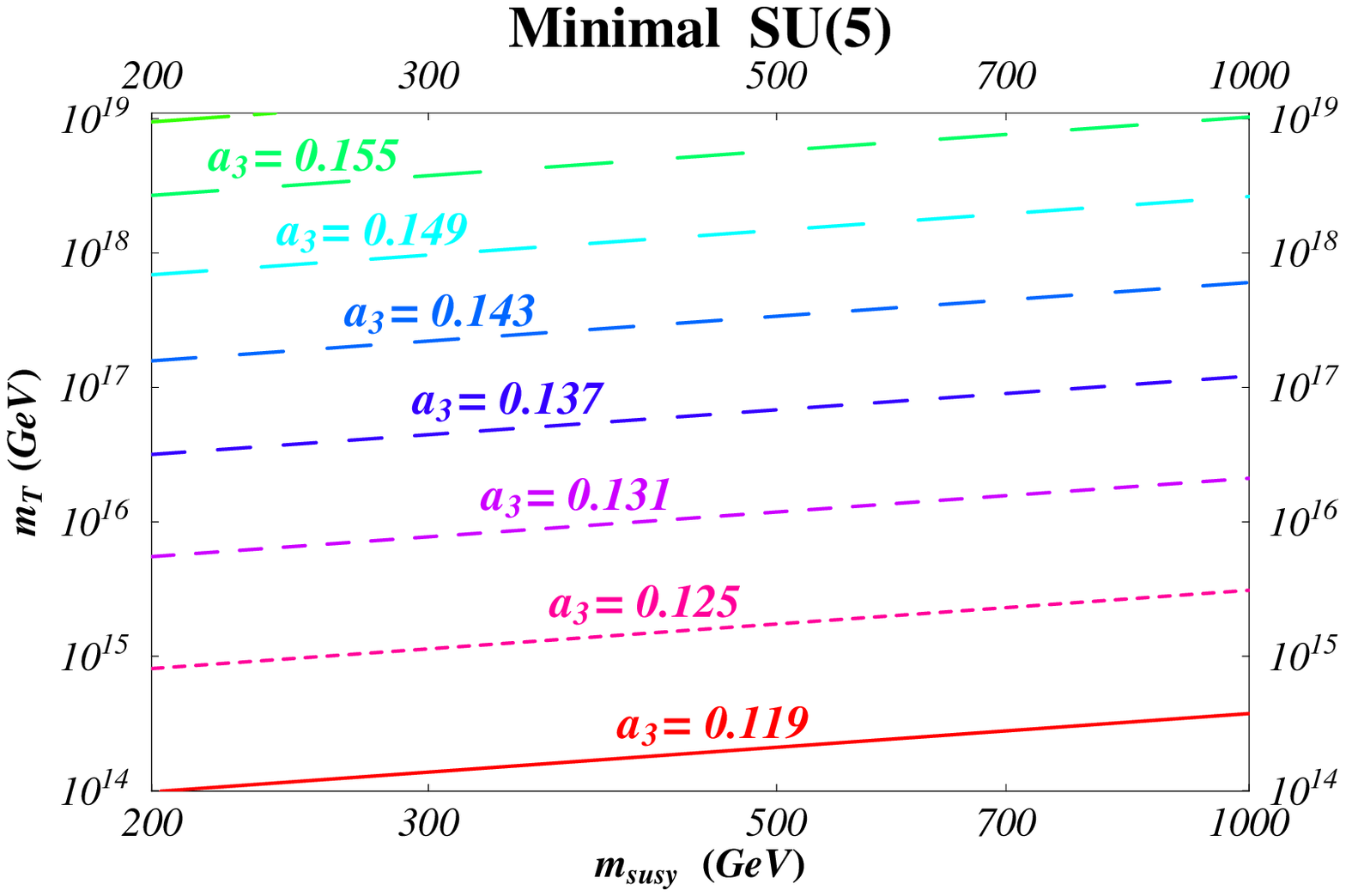,width=1.1\textwidth}}
\end{figure}

\begin{figure}[p]
\centerline{\psfig{file=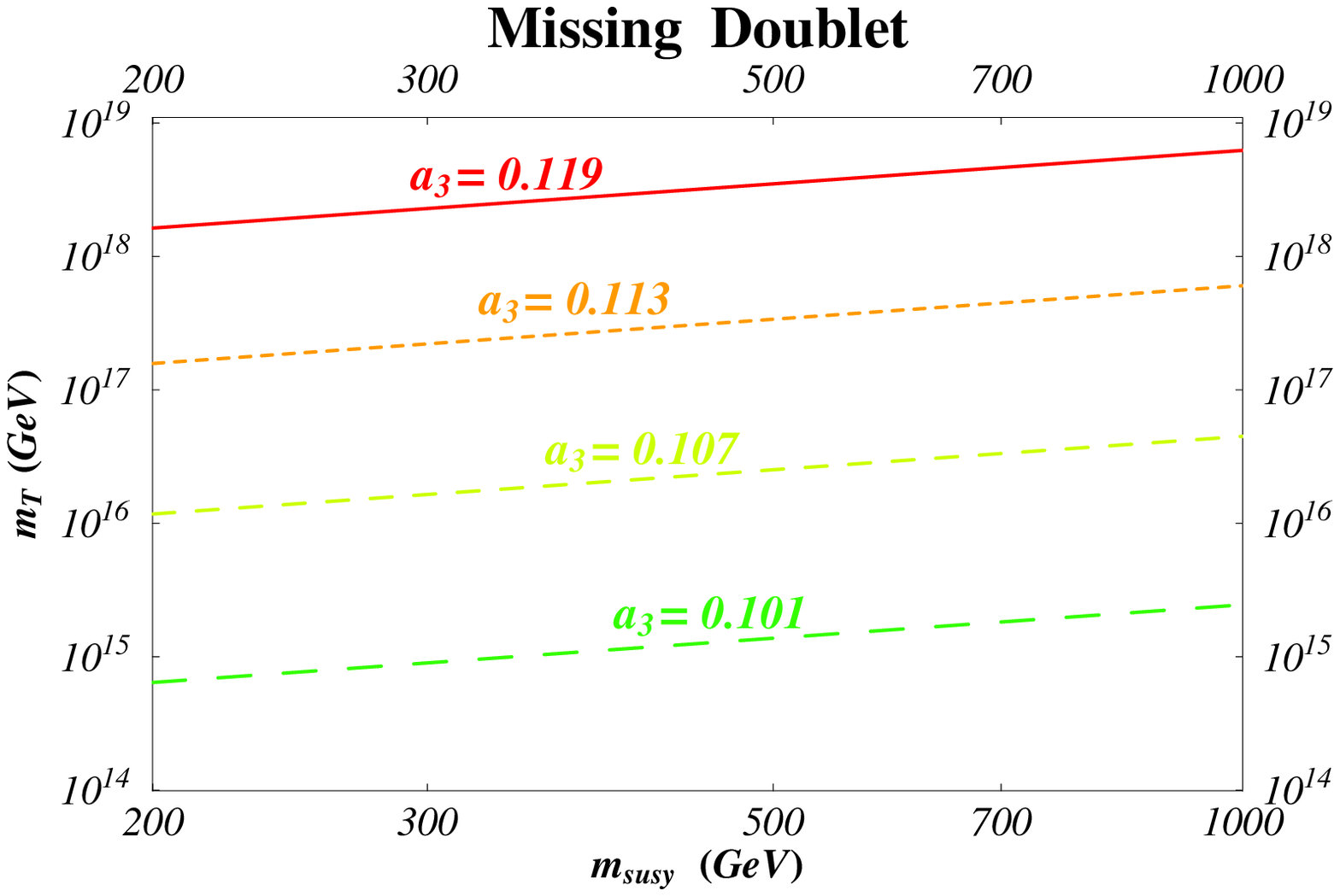,width=1.1\textwidth}}
\caption{Contours of $a_{3}\equiv \alpha_{s}(m_Z)$ in the plane $(m_{SUSY},m_T)$, 
for the minimal $SU(5)$ and the missing doublet
model. The SUSY spectrum is parametrized as in Table 1.}
\label{alpha3}
\end{figure}

Clearly, there is an uncertainty on $k(SUSY)$, related to the possibility of varying
both the parametrization used and 
the relations assumed between the parameters $m_0$, $m_{1/2}$, $\mu$, $m_H$.
An overall uncertainty of $\pm 0.005$ on $a_3$ has been estimated in 
ref. \cite{lp} by
scanning many models consistent with electroweak symmetry breaking, 
a neutral lightest supersymmetric particle and sparticle masses above the experimental 
bound and below $\sim$ 2 TeV 
%This corresponds to an uncertainty on 
%$k(SUSY)+1/(2\pi)\log(m_{SUSY}/m_Z)$ of about $\pm 0.35$ 
\footnote{
We stress that in the present analysis we are assuming universal soft breaking 
parameters at the cut-off scale. If we relax this assumption, then a larger range for
$a_3$ could be obtained. Indeed $a_3$ is particularly sensitive to
wino and gluino masses, that, at the electroweak scale, 
are approximately in the ratio 1:3 
when a universal boundary condition on gaugino masses is imposed. It has been observed 
that, by inverting this ratio, a negative contribution of about $-0.01$ to
$a_3$ is obtained \cite{rs}.}.

Another source of uncertainty on $a_3$ is related to the unknown physics above the cut-off scale of the
grand unified theory. There can be threshold effects due to new heavy particles or even non perturbative effects
that arise in the underlying fundamental theory. These effects can be estimated from the non-renormalizable
operators, suppressed by $\langle Y\rangle/\Lambda$, that split the gauge couplings 
at the scale $M_{GUT}^{(0)}$ \cite{nro}.  
For generic, order one coefficients and barring cancellations among different terms, the presence of these
operators may affect $a_3$ by additional contributions of about $\pm 0.005 M_{Pl}/\Lambda$ \cite{lp}. As we will see
the model under discussion requires $\Lambda\approx 0.1 M_{Pl}$. Therefore, 
to maintain the good agreement
between the experimental and predicted values of $a_3$, we need a suppression 
of this contribution 
by about a factor of 10, which we do not consider too unnatural. 

In view of the large theoretical uncertainties on $a_3$ we cannot firmly
conclude that the gauge coupling unification fails in the minimal model
while it is completely successful in the realistic one.
However we find very encouraging that, by solving
the doublet-triplet splitting problem within the missing partner model,
acceptable values of $a_3$ can be easily obtained and that they
are directly related to a proton lifetime potentially
larger than in the minimal model.

Due to the large matter content the model is not asymptotically
free and the gauge coupling constant $\alpha_5$ blows up at
the scale
\beq
M_{GUT}~~ {\rm exp}{\dd\left(\frac{\pi}{26 a_5}\right)}~~~~~.
\label{blow}
\eeq
Taking into account that threshold effects modify $a_5$ with respect
the LO value $a_{5}^{(0)}\approx 1/24$, we find that the pole occurs
near $10^{17}$ GeV, the precise value depending on the details
of the heavy spectrum.

\section{Yukawa couplings}

To reproduce the fermion mass spectrum we must further 
specialize the Yukawa couplings
introduced in eq. (\ref{w3}):
\bea
w_3 &=& \frac{1}{4}  \Psi_{10}^{ab}~ G_u~ \Psi_{10}^{cd}~ 
H^e \epsilon_{abcde}
+ \frac{1}{4}  \Psi_{10}^{ab}~ G_{\overline{50}}~ \Psi_{10}^{cd}~ 
{H_{\overline{50}}}_{abcd}
\nn\\
&+& \sqrt{2}~ \Psi_{10}^{ab}~ G_d~ {\Psi_{\bar 5}}_a~ {\bar H}_b
+\frac{1}{\Lambda}~\sqrt{2}~ \Psi_{10}^{ab}~ F_d~ {\Psi_{\bar 5}}_c~ {\bar H}_d~ Y^{cd}_{ab}\nn\\
&+& \Psi_1~  G_\nu~ {\Psi_{\bar 5}}_a~ H^a
- \frac{1}{2}~ M~ \Psi_1~ G_M~ \Psi_1+...
\eea
where $G_r~(r=u,d,\nu,M,{\overline{50}})$ is proportional to $G_r(\langle X\rangle,0)$ of eq. (\ref{w3}). We have
explicitly introduced a term linear in $Y$, whose couplings are described
by the $\langle X\rangle$-dependent matrix $F_d$.
There are other terms linear in $Y/\Lambda$, not explicitly
given above. In particular we may insert $Y/\Lambda$ in the renormalizable
term providing masses to the up type quarks. We neglect such a term 
since, on the one hand, the matrix $G_u$ is already sufficient to
correctly reproduce the up quark masses and, on the other hand,
this operator would not significantly modify the results for proton decay. 
As we will show, $Y/\Lambda$ is close to 0.1 in our model and 
higher order terms in the $Y/\Lambda$ expansion can be safely neglected.
The interaction term involving  ${H_{\overline{50}}}_{abcd}$ does 
not contribute
to the fermion spectrum, but it will be relevant for the proton 
decay amplitudes. 

The term linear in $Y$ in the previous equation is sufficient to 
differentiate the spectra in the charged lepton and down quark sectors.
We get the following Dirac mass matrices:
\beq
m_{u,\nu}=y_{u,\nu}~ \frac{v_u}{\sqrt{2}}~~~,~~~~~m_{d,e}=y_{d,e}~ 
\frac{v_d}{\sqrt{2}}~~~~~,
\eeq
\beq
\begin{array}{lll}
y_u=G_u~~~& &y_\nu=G_\nu~~~~~,\\
& \\
y_d=G_d + \dd\frac{\langle Y\rangle}{\Lambda}F_d~~~& & 
y_e^T=G_d - 3 \dd\frac{\langle Y\rangle}{\Lambda}F_d~~~,
\end{array}
\eeq
where $v_u$, $v_d$ and $\langle Y\rangle$ parametrize the vevs of $H$, ${\bar H}$ and
$Y$ respectively. The fermion spectrum can be easily fitted by appropriately
choosing the numerical values of the matrices $G_u$, $G_d$, $F_d$, $G_\nu$  
and $G_M$. The most general fitting procedure would leave a large number of free 
parameters. Here we limit ourselves to the discussion of one particular example. 
In agreement with eq. (\ref{Mnu}), we take:

\beq
G_u=
\left[\matrix{(-0.51+0.61i) \lambda^6&   (0.42-0.70i) \lambda^5&   (0.27+0.86i) \lambda^3\cr
(0.42-0.70i) \lambda^5&   (-0.39+0.52i) \lambda^4&   (-0.30-1.14i) \lambda^2\cr
(0.27+0.86i) \lambda^3&   (-0.30-1.14i) \lambda^2&           1.39}
\right]~~~,
\label{gu}
\eeq

\beq
G_d=\lambda^4
\left[\matrix{(2.39-1.11i) \lambda^5&   (0.33-0.59i) \lambda^3&   (0.13+0.45i) \lambda^3\cr
(0.87+0.55i) \lambda^4&   (2.76+0.89i) \lambda^2&   (0.69-0.51i) \lambda^2\cr
(-1.50+0.94i) \lambda^2&   (0.45+1.78i)           &            1.94}
\right]~~~,
\label{gd}
\eeq

\beq
\frac{\langle Y\rangle}{\Lambda} F_d=\lambda^4
\left[\matrix{
(0.38-0.18i) \lambda^5&  (-0.16-0.06i) \lambda^3&   (0.07+0.04i)\lambda^3\cr
(-0.08-0.05i) \lambda^4&     (-0.20-0.15i) \lambda^2&    (0.15-0.11i) \lambda^2\cr
(-0.09+0.12i) \lambda^2&     (0.07+0.14i) &               -0.19}        
\right]~~~,
\label{fd}
\eeq 

\beq
G_\nu=
\left[\matrix{
(-0.78-0.19i) \lambda^3&  (0.52-0.34i) \lambda& (1.38+0.39i) \lambda^2\cr 
(-1.23-0.34i)\lambda&      0&          (1.04+1.31i)\cr
(0.45+1.18i)\lambda&  0 &        0.8+1.2i} 
\right]~~~,
\eeq

\beq
G_M=
\left[\matrix{
(1.50+0.55i) \lambda^2&    (1.41+1.19i)&    (0.35-1.53i) \lambda\cr
(1.41+1.19i)&             0&           0\cr            
(0.35-1.53i) \lambda&    0&      1.26+1.48i}
\right]~~~.
\eeq

While the generic pattern of $G_u$, $G_d$, $F_d$, $G_\nu$ and $G_M$ is dictated by
the $U(1)$ flavour symmetry, the precise values of the coefficients multiplying
the powers of $\lambda$ are chosen to reproduce the data. 
We take $\lambda\equiv\langle X\rangle/\Lambda=0.25$, $\tan\beta=1.5$ and $M=0.9\cdot 10^{15}$ GeV.    
In SU(5) the matrix $G_u$ contains two additional phases \cite{ncafasi},
$\phi_1$ and $\phi_2$ that have been set to zero in eq. (\ref{gu}).
These phases do not affect the fermion spectrum but enter the
proton decay amplitude. When discussing the proton decay we will
analyze also the dependence on $\phi_1$ and $\phi_2$.

From the above matrices we obtain, at the unification scale:

\beq
\begin{array}{lll}
m_t=200~ {\rm GeV}& m_c=0.27~ {\rm GeV}& m_u=0.9~ {\rm MeV}\\
   m_b=1.0~ {\rm GeV}& m_s=26~   {\rm MeV}& m_d=1.1~ {\rm MeV}\\
m_\tau=1.1~ {\rm GeV}& m_\mu=71~  {\rm MeV}& m_e=0.34~ {\rm MeV}
\end{array}
\eeq
\beq
|V_{us}|=0.22~~~~~|V_{ub}|=0.0022~~~~~ |V_{cb}|=0.052~~~~~J=1.9\cdot 10^{-5}~~~,
\eeq 
where $J$ is the CP-violating Jarlskog invariant.
In the neutrino sector, we find:
\beq
m_1=0.81\cdot 10^{-3} {\rm eV}~,~~~~m_2= 0.88\cdot 10^{-3} {\rm eV}~,
~~~~m_3=0.061~ {\rm eV}~~~.
\eeq
More precisely:
\beq
\Delta m^2_{sol}\equiv m_2^2-m_1^2=1.1\cdot 10^{-7}~ {\rm eV}^2~~,~~~~~
\Delta m^2_{atm}\equiv m_3^2-m_2^2=3.7\cdot 10^{-3}~ {\rm eV}^2~~~. 
\eeq
The neutrino mixing angles are
\beq
\theta_{12}\sim \frac{\pi}{4}~~,~~~~~\theta_{23}\sim \frac{\pi}{4}~~,~~~~~\theta_{13}=0.06~~~.
\eeq
%We have exploited here the well known fact that the operators obtained via an insertion
%of $Y$ differentiate between the down quarks and the charged leptons and may be arranged
%to break in the desired way the rigid mass relation of minimal $SU(5)$.
Until now we have not specified the value of $\langle Y\rangle/\Lambda$. We know 
that $\langle Y\rangle$
should be
around $M_{GUT}^{(0)}$. The cut-off $\Lambda$ cannot be too close
to $\langle Y\rangle$, otherwise most of the spectrum of the model would lie beyond the cut-off.
At the same time $\Lambda$ cannot be too large: it is bounded from above
by the scale at which the $SU(5)$ gauge coupling blows up, which, as we 
see from eq. (\ref{blow}), occurs more or less one order of magnitude below the Planck mass. This is
welcome. If, for instance, we take $\langle Y\rangle/\Lambda=0.05-0.1$, it is reasonable to neglect 
multiple insertions of $Y$ in the Yukawa operators. At the same time, in the example given in 
eq. (\ref{fd}), 
the coefficients of the powers of $\lambda$ in $F_d$ remain of order one, even if  
$\langle Y\rangle/\Lambda$ is as small as 0.05.
A too large cut-off would have been ineffective in separating down quarks from charged
leptons unless we had chosen unnaturally large coefficients in the allowed operators.   
What is usually considered a bad feature of the missing partner mechanism -
the lack of perturbativity before the Planck mass - turns out here to be an advantage
to provide a correct description of the fermion spectrum.

The neutrino sector is quite similar to one of the two options described in 
ref. \cite{af23}.
We obtain a bimaximal neutrino mixing with the so-called 
LOW solution to the solar
neutrino problem  \footnote{Latest preliminary results from Super-Kamiokande
\cite{lastSK}, 
including constraints from day-night spectra, seem in fact 
to prefer bimaximal neutrino mixing and to revamp interest 
in the LOW solution.}. Within the same $U(1)$ flavour symmetry 
considered here we could as well
reproduce the vacuum oscillation solution, by appropriately tuning the 
order one
coefficients in $G_\nu$ and $G_M$. We recall that the value 
of $M$ required to fit
the observed atmospheric oscillations is probably 
somewhat small in the context of
$SU(5)$, where a larger scale, closer to the cut-off 
$\Lambda$, is expected.
This feature might be improved by embedding the model in 
$SO(10)$ where $M$
is directly related to the $B-L$ breaking scale.

In conclusion, the known fermion spectrum can be reproduced starting from a superpotential
with order one dimensionless coefficients. Mass matrix elements for charged leptons and down quarks
match only within $10-20\%$, due to $\langle Y\rangle/\Lambda\approx 0.1$, and this produces the required 
difference between the two sectors. The neutrino mixing is necessarily bimaximal in our model,
with either the LOW or the vacuum oscillation solution to the solar neutrino problem.

\section{Proton Decay}

Similarly to the case of minimal $SU(5)$, we expect that the 
main contribution to proton decay comes from  
the dimension five operators \cite{five} originating at the 
grand unified scale when the colour triplet superfields
are integrated out \cite{enr, susygut, hmy}. 
We denote the colour triplets contained in $H$, ${\bar H}$, $H_{50}$, 
${H_{\overline{50}}}$ 
by $H_{3u}$, $H_{3d}$, $H'_{3d}$ and $H'_{3u}$, respectively. 
The part of the superpotential depending
on these superfields reads:
\bea
w&=&H_{3u}\left[-\frac{1}{2} Q G_u Q + U^c G_u E^c\right]\nn\\
&+&H_{3d}\left[Q\hat{C} L 
+ U^c \hat{D} D^c\right]\nn\\
&+&H'_{3u}\left[-\frac{1}{2} Q G_{\overline{50}} Q + U^c G_{\overline{50}} E^c\right]\nn\\
&+&c_2 \langle Y\rangle~ H'_{3d} H_{3u} + c_3 \langle Y\rangle~ H_{3d} H'_{3u} + c_4 \langle X\rangle~ H'_{3d} H'_{3u} +...
\eea
where
\beq
\hat{C}=-G_d-\frac{\langle Y\rangle}{\Lambda} F_d~~,~~~~~~~~
\hat{D}=G_d-\frac{\langle Y\rangle}{\Lambda} F_d
\eeq
and $Q$, $L$, $U^c$, $D^c$ and $E^c$ denote as usual the chiral multiplets associated to 
the three fermion generations.
Notice that, at variance with the minimal $SU(5)$ model, an additional interaction term
depending on $H'_{3u}$ is present.
%The coupling constants in $G_{\overline{50}}$ are completely independent from 
%those in $G_u$, even though both are constrained by the same $U(1)$ flavour symmetry.
By integrating out the colour triplets we obtain the following effective superpotential:
\beq
w_{eff}=\frac{1}{m_T}\left[Q \hat{A} Q~
Q \hat{C} L + U^c \hat{B} E^c~
U^c \hat{D} D^c\right]+...
\eeq
where $m_T$ has been defined in eq. (\ref{mT}),
\beq
\hat{B}=-2 \hat{A}=\left(G_u-\frac{c_2}{c_4}\frac{\langle Y\rangle}{\langle X\rangle} G_{\overline{50}}\right)
\eeq
and dots stand for terms that do not violate baryon or lepton number.
Minimal $SU(5)$ is recovered by setting $G_{\overline{50}}=F_d=0$. 
In that case the matrices 
$\hat{A}$, $\hat{B}$, $\hat{C}$ and $\hat{D}$ 
are determined by $G_u$ (in which now also $\phi_1$ and $\phi_2$ play a role) 
and $G_d$ and therefore strictly related 
to the fermionic spectrum \cite{enr}.
In our case we have a distortion due to the terms proportional to $G_{\overline{50}}$ and $F_d$.
These distortions have different physical origins. On the one hand the terms containing $F_d$ are 
required to avoid the rigid mass relation of minimal $SU(5)$. They are suppressed by
$\langle Y\rangle/\Lambda$ and we expect a mild effect from them \cite{inserz}.
On the other hand the terms containing $G_{\overline{50}}$ are a consequence of the missing doublet mechanism
and they might be as important as those of the minimal model.

At lower scales the dimension five operators give rise to the four fermion operators
relevant to proton decay, via a ``dressing'' mainly due to 
chargino exchange \cite{five,enr,dress}. 
When considering the operators $QQQL$ the main contribution, 
here called $L_1$,
comes from the exchange of a wino 
\footnote{Gluino dressing contributions cancel among each other in case of
degeneracy between first two generations of squarks \cite{ehnt,gluino}.}. 
Charged higgsino exchange provides instead
the most important dressing of the operators $U^c U^c D^c E^c$~\cite{goto}. We term $L_2$ the 
leading four-fermion operator in this case. Beyond $L_1$ and $L_2$ other 6 operators are generated
by chargino exchange, 3 from $QQQL$ and 3 from $U^c U^c D^c E^c$. The contributions
to the proton decay amplitudes from these operators are suppressed by at least 
$\lambda^2$ with respect to those associated to $L_1$ and $L_2$, and can be safely neglected
in the present estimate. $L_1$ and $L_2$ are given by:
%\bea
%L_1&=&\frac{K_1}{m_D}\frac{g^2}{8 \pi^2}~[f(\tilde{u},\tilde{d})+f(\tilde{u},\tilde{e})]~ 
%(\nu_l d_i)~(d_j u_k)~ P_{ij}~ Q_{kl}+{\rm h.c.}\nn\\
%L_2&=&-\frac{K_2}{m_D}\frac{1}{8 \pi^2}~f(\tilde{u^c},\tilde{e^c})~ 
%(\bar{d}_k \bar{\nu}_l)~(u^c_i d^c_j)~ S_{kl}~ T_{ij}+{\rm h.c.}~~~,
%\eea
\bea
L_1&=&C^1_{ijkl}~ (\nu_l d_i)~(d_j u_k)+{\rm h.c.}\nn\\
L_2&=&C^2_{ijkl}~ (d_i \nu_j)~(\overline{u^c}_k \overline{d^c}_l)+{\rm h.c.}~~~,
\eea
where $i,j,k,l$ are generation indices and
\bea
C^1_{ijkl}&=&\frac{K_1}{m_T}\frac{g^2}{8 \pi^2}~[f_{\tilde{w}}(\tilde{u},\tilde{d})+f_{\tilde{w}}(\tilde{u},\tilde{e})]~
P_{ij}~ Q_{kl}\nn\\
C^2_{ijkl}&=&-\frac{K_2}{m_T}\frac{1}{8 \pi^2}~f_{\tilde{h}}(\tilde{u^c},\tilde{e^c})
S_{ij}^*~ T_{kl}^*
\eea
\beq
\begin{array}{lll} 
P=2~ L_d^T \hat{A}~ L_d & &Q=L_u^T \hat{C}~ L_e\\
& \\
S=L_d^\dagger~ y_u^*~ \hat{B}~ y_e^\dagger~ L_e^*& &T=R_u^*~ \hat{D}~ R_d^\dagger~~~.
\end{array}
\eeq
$L_{u,d,e}$ and $R_{u,d,e}$ are the unitary matrices that diagonalize the fermion mass matrices:
\beq
L_u^T    y_u R_u^\dagger = (y_u)_{diag}~~~~~~~~- L_d^T y_d R_d^\dagger = (y_d)_{diag}~~~~~~~~
- L_e^T y_e R_e^\dagger = (y_e)_{diag}~~~,
\eeq
with $(y_f)_{diag}~~(f=u,d,e)$ diagonal and positive. The quark mixing matrix
is $V_{CKM}=L_u^\dagger L_d$. $K_{1,2}$ are constants accounting for the renormalization
of the operators from the grand unification scale down to 
1 GeV \cite{enr, hmy, ren}.
In our estimates we take $K_1=K_2=10$. Finally, $f$ is a function coming 
from the loop integration:
\beq
f_{\tilde{c}}({\tilde{a}},{\tilde{b}})=\frac{m_{\tilde{c}}}{2}
\frac{1}{(m^2_{\tilde{a}}-m^2_{\tilde{b}})}\left[
\frac{m^2_{\tilde{a}}}{m^2_{\tilde{a}}-m^2_{\tilde{c}}}\log
\frac{m^2_{\tilde{a}}}{m^2_{\tilde{c}}}-
\frac{m^2_{\tilde{b}}}{m^2_{\tilde{b}}-m^2_{\tilde{c}}}\log
\frac{m^2_{\tilde{b}}}{m^2_{\tilde{c}}}\right]~~~.
\eeq
We parametrize the partial rates 
according to the results of a chiral lagrangian computation \cite{chiral}.
The rates for the dominant channels are given in Table 2.

To estimate the proton decay rates we should specify some important parameters.
First of all the mass $m_T$. From the discussion of the threshold 
corrections we know that a large value for $m_T$ is preferred in our model.
We should however check that this value can be obtained with reasonable choices
of the parameters at our disposal. The unification conditions constrain $\langle Y\rangle$
in a small range around $10^{16}$ GeV \footnote{More precisely, after the inclusion
of threshold corrections and two loop effects, the last equality in eq. (\ref{17})
becomes a relation among the masses of the super-heavy particles and the 
leading order quantity $M_{GUT}^{(0)}$. This relation limits the allowed range
for the heavy gauge vector bosons and indirectly pushes the $SU(5)$
breaking vev $\langle Y\rangle$ close to $10^{16}$.}. Then the cut-off
scale and the $X$ vev are fixed by the phenomenological requirements
$\langle Y\rangle\approx 0.05~ \Lambda$ and $\lambda\equiv \langle X\rangle/\Lambda=0.25$.
Large values of $m_T$ could be obtained either by taking $m_{T_1} m_{T_2}$ large
or by choosing a small $m_\phi$. This last possibility is however not practicable,
since the gauge coupling $\alpha_5$ blows up at approximately 20 $m_\phi$:
it is not reasonable to push $m_\phi$ below $5\cdot 10^{15}$ GeV. 
\\[0.1cm]
%\begin{table}
%\tcaption{Proton decay rates}
{\begin{center}
\footnotesize
\begin{tabular}{|c|c|}   
\hline
& \\                         
channel & rate\\ 
& \\
\hline
& \\
$K^+ \bar{\nu}_\mu$ & $X_K \left\vert \beta \left[\frac{2}{3}\frac{m_p}{m_B} D C^1_{1212}+
\left(1+\frac{m_p}{3 m_B}(D+3F)\right)C^1_{2112}\right]\right\vert ^2$\\ 
& \\
\hline  
& \\
$K^+ \bar{\nu}_\tau$ & $X_K \left\vert \beta \left[\frac{2}{3}\frac{m_p}{m_B} D C^1_{1213}+
\left(1+\frac{m_p}{3 m_B}(D+3F)\right)C^1_{2113}\right]\right.$\\
& \\
&$\left.+\alpha \left[\frac{2}{3}\frac{m_p}{m_B} D C^2_{1312}+
\left(1+\frac{m_p}{3 m_B}(D+3F)\right)C^2_{2311}\right]\right\vert^2$\\ 
& \\
\hline
& \\
$\pi^+ \bar{\nu}_\mu$ & $X_\pi\left\vert\beta \left(1+D+F\right) C^1_{1112}\right\vert^2$\\
& \\
\hline
& \\
$\pi^+ \bar{\nu}_\tau$ & $X_\pi\left\vert\beta \left(1+D+F\right) C^1_{1113}
+\alpha \left(1+D+F\right) C^2_{1311}\right\vert^2$\\
& \\
\hline
\end{tabular} 
\end{center}}
\vspace{3mm}
Table 2. Proton decay rates. We define $X_{K,\pi}=(m^2_p-m^2_{K,\pi})^2/(32\pi m_p^3 f_\pi^2)$; 
$m_p$, $m_K$, $m_\pi$ are the proton, $K^+$ and $\pi^+$ masses;
$m_B$ is an average baryon mass, $f_\pi$ is the pion decay constant;
$D$ and $F$ are coupling constants between baryons and mesons in the relevant
chiral lagrangian; $\beta$ and $\alpha$ parametrize the hadronic matrix element. 
In our estimates we take $m_p=0.938$ GeV, $m_B=1.150$ GeV, $m_K=0.494$ GeV, $m_\pi=0.140$ GeV, 
$f_\pi=0.139$ GeV, $D=0.8$, $F=0.45$ \cite{DF}, $\beta=-\alpha=0.003$ GeV$^3$ \cite{albe}. 

The only remaining freedom to obtain the desired large value for $m_T$
is represented by the coefficients $c_2$ and $c_3$ that, however,
cannot be taken arbitrarily large. We should also check that
all the heavy spectrum remains below $\Lambda$ and this requirement
imposes a further constraint on our parameters.
A choice that respects all these requirements
is provided by
\footnote{It is clear that the extreme values of the coefficients $c_2$ and $c_3$ here adopted 
raise doubts on the validity of the perturbative approach that has been exploited in several aspects 
of the present analysis.
We adhere to this extreme choice also to show the difficulty met to obtain acceptable
phenomenological results within a not too complicated scheme.}:
\beq
c_1=0.004 ~~~,~~~~~c_2=c_3=19.3,~~~~~c_4=0.7~~~,
\eeq
\beq
\langle X\rangle=3.0\cdot 10^{16}~{\rm GeV},~~~~~\langle Y\rangle=5.7\cdot 10^{15}~{\rm GeV}~~~~~
(\Lambda=1.2\cdot 10^{17}~{\rm GeV})~~~,\eeq
which leads to:
\beq
M_{GUT}=2.9\cdot 10^{16}~{\rm GeV}~~~,~~~~~m_\phi=2.0\cdot 10^{16}~{\rm GeV}~~~,
\eeq
\beq
%m_\phi=2.0\cdot 10^{16}~{\rm GeV}~~~,~~~~~~
m_{T_1}=1.2\cdot 10^{17}~{\rm GeV}~~~,~~~~~
m_{T_2}=1.0\cdot 10^{17}~{\rm GeV}~~~,~~~~~m_T=6\cdot 10^{17}~{\rm GeV}~~~.
\eeq
The heavy sector of the particle spectrum is displayed in fig. 2. With the above values we also obtain:
\beq
\frac{c_2}{c_4}\frac{\langle Y\rangle} {\langle X\rangle}\simeq 5.5~~~.
\eeq

To evaluate the loop function $f$
we also need the spectrum of the supersymmetric particles. As an example
we take here the same spectrum considered in table 1, with $m_{SUSY}=250$ GeV.
This leads to a squark mass of about $800$ GeV, a slepton mass of approximately $350$ GeV
a wino mass of 250 GeV and a charged higgsino mass of 125 GeV.

The matrix ${G_{\overline{50}}}$ is not directly related to any accessible observable 
quantity and the only constraint we have on it comes from the $U(1)$ flavour symmetry
that requires the following general pattern:
\beq
{G_{\overline{50}}}=
\left[\matrix{\lambda^7&   \lambda^6&   \lambda^4\cr
\lambda^6&   \lambda^5&   \lambda^3\cr
\lambda^4&   \lambda^3&   \lambda}
\right]~~~.
\label{g50b}
\eeq

\begin{figure} [t]
\centerline{
\psfig{file=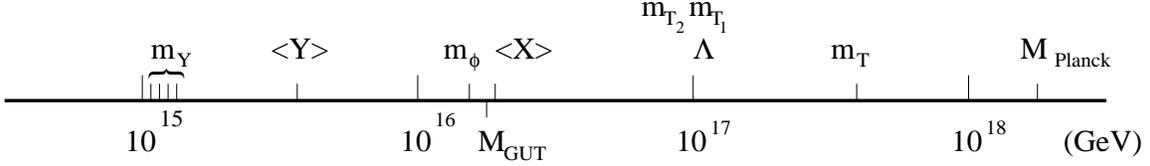,width=0.90\textwidth}
}
\caption{Heavy sector of the spectrum.}
\label{spectrum}
\end{figure}

The texture for ${G_{\overline{50}}}$ has an overall suppression factor $\lambda$ 
compared to $G_u$. Therefore the contribution of ${G_{\overline{50}}}$ to the matrices
$\hat{A}$ and $\hat{B}$ is comparable or even slightly larger than the minimal
contribution provided by $G_u$. The interference between the amplitude with $G_u$ and the one 
with ${G_{\overline{50}}}$ can be either constructive or destructive, depending
on the relative phases between the two terms.
We have scanned several examples for ${G_{\overline{50}}}$, obtained by
generating random coefficients for the order one variables in eq. (\ref{g50b}).
By keeping fixed all the remaining parameters we obtain a proton decay 
rate 
% o lifetime?%
in the range $8\cdot 10^{31}-3\cdot 10^{34}$ ys for the channel 
$p\to K^+ \overline{\nu}$ and a rate
between $2\cdot 10^{32}$ ys and $8\cdot 10^{34}$ ys for the channel 
$ p\to\pi^+ \overline{\nu}$.
For comparison, considering the same choice of parameters but setting 
${G_{\overline{50}}}=F_d=0$, 
we obtain $9\cdot 10^{32}$ ys and $2 \cdot 10^{33}$ ys respectively, for the above 
channels. The present 90\% CL bound on $\tau/BR(p\to K^+\bar{\nu})$ is $1.9~10^{33}$ ys \cite{lastSK}.
These estimates have been obtained by setting to zero the two physical phases $\phi_1$, $\phi_2$
contained in the matrix $G_u$. These additional parameters may increase the uncertainty
on the proton lifetime. For instance, in minimal SU(5), the proton decay rates for the
channels considered above, change by about one order of magnitude when $\phi_1$ and $\phi_2$
are freely varied between 0 and $2 \pi$. 
Even when the inverse decay rates for the channels $K^+ \bar{\nu}$ and $\pi^+ \bar{\nu}$
are as large as $10^{34}$ ys, they remain the dominant contribution to the proton lifetime.
Indeed, since the heavy vector boson mass, $M_{GUT}$, is equal to $2.9~ 10^{16}$ GeV in our model,
the dimension 6 operators provide an inverse decay rate for the channel $e^+ \pi^0$ larger than 
$10^{36}$ ys. 

The effective theory considered here breaks down at the cut-off 
scale $\Lambda$. We expect 
additional non-renormalizable operators contributing 
to proton decay amplitudes
from the physics above the cut-off. By assuming dimensionless coupling constants of order one,
in unified models without flavour symmetries the proton lifetime induced by these operators
is unacceptably short, even when $\Lambda=M_{Pl}$ \cite{ehnt,nroforp}. 
In our case these contributions are adequately suppressed 
by the $U(1)$ symmetry. 
If we compare the amplitude $A_{nr}$ induced by the new non-renormalizable operators with the 
amplitude $A_3$ coming from the triplet exchange, we obtain, for the generic decay channel,
\beq
\frac{A_{nr}}{A_3}\approx \frac{c_2 c_3}{c_4} \left(\frac{\langle Y\rangle}{\Lambda}\right)^2\approx 1~~~~~.
\eeq
This supports the conclusion that a proton lifetime range considerably 
larger than the one estimated in minimal models is expected in our case.

\section{Conclusions}

We have constructed an example of SUSY $SU(5)$ GUT model, with an additional
$U(1)$ flavour symmetry, which is not plagued by the
need of large amounts of fine tunings, like those associated with
doublet-triplet splitting in the minimal model, and leads
to an acceptable phenomenology. This includes coupling unification with a
value of $\alpha_s(m_Z)$ in much better agreement
with the data than in the minimal version, an acceptable pattern for
fermion masses and mixing angles, also including
neutrino masses and mixings, and the possibility of a slower proton decay
than in the minimal version, compatible with the
present limits (in particular the limit from Super-Kamiokande of about
$2\cdot 10^{33}$ ys for the channel $p\rightarrow K^+
\bar{\nu}$). In the neutrino sector the present model is a special case of
the class of theories discussed in ref. \cite{af23}. The
preferred solution in this case is one with nearly maximal mixing both for
atmospheric and solar neutrinos. The $U(1)$ flavour symmetry
plays a crucial role by protecting the light doublet Higgs mass from
receiving large mass contributions from higher dimension
operators and by determining the observed hierarchy of fermion masses and
mixings. Of course, the $U(1)$ symmetry can only
reproduce the order of magnitude of masses and mixings, while more
quantitative relations among masses and mixings can only
arise from a non abelian flavour symmetry.  

A remarkable feature of the model is that the presence of the
representations
$50$,
$\overline{50}$ and
$75$, demanded by the missing partner mechanism for the solution of the
doublet-triplet splitting problem, directly
produces, through threshold corrections at $M_{GUT}$ from the $75$, a
decrease of the value of $\alpha_s(m_Z)$ that  
corresponds to coupling unification and an increase of the effective mass
that mediates proton decay by a factor of
typically 20-30.  As a consequence the value of the strong coupling is in
better agreement with the experimental value and
the proton decay rate is smaller by a factor 400-1000 than in the minimal
model. The presence of these large representations
also has the consequence that the asymptotic freedom of $SU(5)$ is spoiled
and the associated gauge coupling becomes non
perturbative below 
$M_{Pl}$. We argue that this property far from being unacceptable can
actually be useful to obtain better results for fermion
masses and proton decay. 

Clearly such a model is not unique: our version is the simplest realistic
model that we could construct. We think it is
interesting because it proves that a SUSY $SU(5)$ GUT is not excluded and
offers a benchmark for comparison with experiment.
For example, even including all possible uncertainties, it is difficult in
this class of models to avoid the conclusion that
proton decay must occur with a rate which is only a factor 10-50 from the
present bounds. Failure to observe such a signal
would require some additional specific mechanism in order to further
suppress the decay rate. Finally it is a generic
feature of realistic models that the region between $M_{GUT}$ and $M_{Pl}$
becomes populated by many states with
different thresholds and also non perturbative phenomena occur. This
suggests that the reality can be more complicated than
the neat separation between the GUT and the string regime which is
postulated in the simplest toy models of GUT's.

\noindent
{\bf Acknowledgements}

\noindent
We would like to thank Zurab Berezhiani, Andrea Brignole, Franco Buccella, Antonio Masiero, 
Jogesh Pati, Antonio Riotto, Anna Rossi, Carlos Savoy and Fabio Zwirner for useful discussions.

\end{document}